\documentstyle[12pt,epsf]{article}
\catcode`\@=11 \@addtoreset{equation}{section}  %%
\renewcommand{\theequation}                     %%
         {\arabic{section}.\arabic{equation}}   %%
\title{The non-Abelian State-Dependent Gauge Field in Optics}
\author{P. Leifer}
\date{ Cathedra of Informatics, Crimea State Engineering and Pedagogical University, \\ 
21 Sevastopolskaya st., 95015 Simferopol, Crimea, Ukraine} 
\begin{document}
\maketitle
\begin{abstract}
The appearance of the geometrodynamical version of the Pancharantam phase factor in the polarization state of the monochromatic beam was systematically discarded during the topological phases measurements has been shown. It means that not only topological structure of the transformation manifolds has the physical significance but the metric structure of the coherent state space of its realization too. The comparison of the local dynamical variables over such states is defined through the parallel transport agrees with the Fubini-Study metric. This parallel transport might be actual in the Qubit coding-encoding processing of the quantum information.  
\end{abstract}

\vskip 0.5cm

The topological character of the phases Berry \cite{Berry1,Berry2}, Aharonov-Anandan \cite{AA1,AA2} and Wilczek -Zee \cite{WZ} arises as a macroscopic environment reaction on the quantum dynamics of an ``immersed" quantum system. The anholonomies of the ``parallel transport" of the state vector are expressed 
as some effective gauge fields reflecting the topological character of the transformation groups of orientations of  macroscopic  elements (polarizers, $\lambda /4$ plates, etc.) of the quantum setup. Therefore it is not so strange that there are close classical analogies of the topological phases in classical physics (e.g. Hannay angle \cite{Berry2}). This is the reason why a dynamic phase should be discarded in order to get definite geometric (topological) phase. Therefore in general it is impossible of course to endow these gauge fields by some fundamental sense. But such gauge fields may by really fundamental in two important cases being appear into the complex projective state space $CP(N-1)$. Firstly, since we believe that rays of quantum states are the fundamental notions at any level. 
Secondly, $CP(1)$ may be treated as the Qubit coherent state space
under the quantum information processing.
In these cases arises a new geometrodynamics phase relates to the affine gauge field. Corresponding gauge fields associated with the curvature of $CP(N-1)$ will be state-dependent and they realize local gauge transformation of the moving quantum frame in $CP(N-1)$ \cite{Le1,Le2,Le3}. They are akin to the Wilczek-Shapere gauge fields related to the problem of a deformable body swim \cite{SW}.
 
Berry's \cite{Berry1,Berry2} and Aharonov-Anandan \cite{AA1,AA2} the ``parallel transport" laws of the quantum state is defined in original Hilbert space. This kind of the parallel transport is not the object of the intrinsic geometry of neither a parameter space (Berry) nor the projective Hilbert state spaces (Aharonov-Anandan); see for example discussion in \cite{AS}. Intention of such definition is to discard dynamical phase shift and to extract pure topological consequences of the rotations of polarizers, $\lambda /4$ plates, etc. However there are some reasons to keep dynamics together with geometry \cite{AA1,Le1,Le3}. In particular the fundamental importance of the complex projective geometry of the state space $CP(N-1)$ \cite{AA2,Hugston1,Jones1,Le1,Le2,Le3} evokes necessity to work in the intrinsic geometry of $CP(N-1)$ close connected with the quantum dynamics.  
 
I will show the geometrodynamics of the light polarization states (the example of the two-level system, N=2) would be restored by the cost of the path-dependent parallel transport in the affine connection agrees with the K\"ahlerian metric (Fubiny-Study metric) in the particular case of the $CP(1)$. The essential differences between my approach and say approach of Anandan and Pati \cite{AP} are firstly, that I use the parallel transport of the local in $CP(1)$ dynamical variables instead of the quantum state transport. Secondly, the geometric frequency I used is local and it is applicable to any superposition state whereas the Anandan-Pati ``reference-section'' of the state is bi-local and it is singular for the orthogonal initial and final states.

I will start with the description of the model setup providing the   unitary evolution of the polarization state of light. The fixed Cartesian reference frame ${(O,x,y,z)}$ in physical space will be used. Initially one has the light beam in the linear polarization state in  $x$-direction $|x>=\frac{1}{\sqrt 2}(|R>+|L>)=\frac{1}{\sqrt 2}(1,1)^T$ propagating along $z$-axes. Then the different polarization states will be expressed as follows:  
$|y>=\frac{1}{\sqrt 2}(|R>-|L>)=\frac{1}{\sqrt 2}(-i,i)^T$,
and then $|R>=(1,0)^T,|L>=(0,1)^T$. The coherent superposition state
will be denoted as usually $|\Psi>=(\Psi^0,\Psi^1)^T$. The Poincar\'e sphere refers to the coordinates ${(o,s_1,s_2,s_3)}$
in iso-space of the polarization. In general the coherence vector lays on the isotropy ``light cone" $s_0^2-s_1^2-s_2^2-s_3^2=0$ where $s_0^2=I^2=<\Psi|\Psi>$ is the square of the beam intensity. It means the coherence vector may ``dive" into the Poincar\'e sphere under non-unitary evolution. 
I will restrict myself by the unitary one. 

The initial state $|x>$ subjected the modulation by the passing through optically active medium (say using the Faraday effect in YIG film magnetizing along the main axes in $z$-direction by the harmonic magnetic field with the frequency $\Omega$ and the amplitude $\beta$). Formally this process may be described by   the unitary matrix action ${\hat h}_{os_3}$
belonging to the isotropy group of the $|R>$ \cite{Le1}.
The coherence vector oscillates along the equator of the Poincar\'e sphere under the modulation having the components
${\vec C}=(<x'|\sigma_1|x'>,<x'|\sigma_2|x'>,<x'|\sigma_3|x'>)=(cos(2 \phi(\Omega t)), sin(2 \phi(\Omega t)),0)$, where $\sigma_{\alpha}$ are Pauli matrices. 
The next step is the dragging with the frequency $\omega$ of the oscillating state $|x'(t)>=\hat{h_{os_3}}|x>$ (in fact the coherence vector ${\vec C}$) up to the ``north pole'' corresponding to the state $|R>$. This may be achieved by the variation of the azimuth of linear polarized state
from $\frac{\theta}{2}=-\frac{\pi}{4}$ up to $\frac{\theta}{2}=\frac{\pi}{4}$
with help the dense flint appropriate length embedded into the sweeping magnetic field. Further this beam should pass the 
$\lambda /4$ plate. This process of variation of the ellipticity
of the polarization ellipse may be described by the unitary matrix
${\hat b}_{os'_1}$
belonging to the coset homogeneous group sub-manifold $SU(2)/S[U(1) \times U(1)]=CP(1)$ of the $|R>$ \cite{Le1}. Without the modulation this dragging leads to the evolution of the initial state along the geodesic of $CP(1)$ and the trace of the coherent vector on the Poincar\'e sphere is the meridian between the equator and the pole. The modulation deforms both the geodesic and the corresponding trace of the tip of the coherence vector leaving on the Poincar\'e sphere during such evolution as it is shown in the Fig.1. 
\vskip .1cm
\begin{figure}[ht]
\centerline{\epsfxsize=\hsize \epsffile{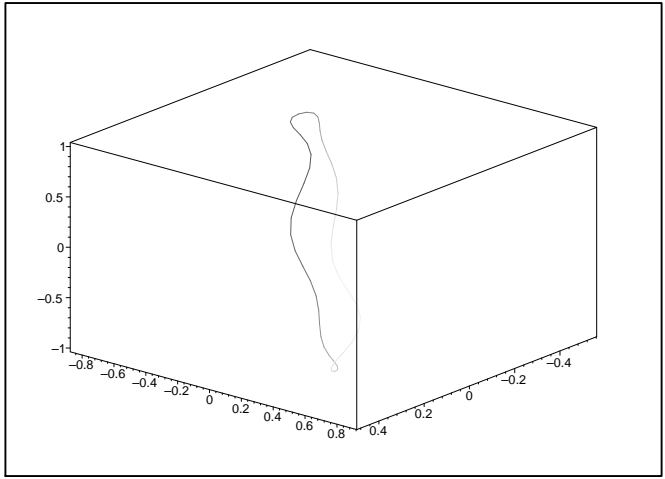}}
\caption{The deformation of the image of the geodesic  on the Poincar\'e sphere during the modulation and the cyclic dragging of the polarization state. The relationship between frequencies is as follows:$\Omega=10 \pi, \omega=\pi$.}
\label{fig.1}
\end{figure} 
\vskip .1cm

The light beam passing through optically active media like YIG film and the dense flint leads to the {\it slow} kinematics of the polarization ellipse shape and its orientation. The fast dynamics
of the electromagnetic potentials or fields corresponding to the dynamical phase may be 
discarded by the transition to the local projective coordinates
\begin{eqnarray}
\pi^1
=\frac{1+\sin(2\eta)}{\cos(2\eta)(\cos(2\Phi)+i\sin(2\Phi)},
\end{eqnarray}
where the ellipticity $\eta$ and inclination (azimuth) $\Phi$ angle may be calculated through the components of the coherence vector. This slow coherent dynamics of the relative amplitudes and phases of the photons may be described by the dispersion law along the optic path $Oz$.

The essentially new element of the coherent dynamics and kinematics is local dynamical variables represented by the tangent vectors to 
$CP(N-1)$ have been introduced \cite{Le1}. Physically interesting
components of the tangent vector fields 
\begin{equation}
\Phi_{\sigma}^i = \lim_{\epsilon \to 0} \epsilon^{-1}
\biggl\{\frac{[\exp(i\epsilon \lambda_{\sigma})]_m^i \Psi^m}{[\exp(i \epsilon
\lambda_{\sigma})]_m^j
\Psi^m }-\frac{\Psi^i}{\Psi^j} \biggr\}=
\lim_{\epsilon \to 0} \epsilon^{-1} \{ \pi^i(\epsilon \lambda_{\sigma}) -\pi^i \}.
\end{equation}
serve as nonlinear representation (realization) of the
$SU(N)$ group action on the arbitrary 
coherent state with the local coordinates
$(\pi^1=\frac{\Psi^1}{\Psi^j},...,\pi^{N-1}=\frac{\Psi^{N-1}}{\Psi^j})$ corresponding to $|\Psi> = \Psi^a |a>$.
The tangent vector fields
$D_{\sigma}=\Phi^i_{\sigma}\frac{\partial}{\partial \pi^i}+c.c.$ 
define the generators of the $SU(N)$ and replace matrices of Pauli in two-level systems with $SU(2)$, Gell-Mann matrices in three-level systems with $SU(3)$ dynamical groups, etc., \cite{Le1}.
In our case these operators give the local {\it polarization  and inversion operators} with the correct commutation relations and provide pure locally unitary reference frame relative Fubini-Study metric. The local approach is close related to the works of Scrotsky and Kusmitchev \cite{SK,S} where systematically used local complex projective coordinate ($\xi(t)$ in their denotations). Furthermore, Scrotsky posed the important question about the character of the cooperative interaction in the ensemble of two-level systems described by the Landau-Lifshiz equation. 

In order to clarify the interaction character one has two realizations of the Riemannian anholonomy. 
   
1. The Riemannian curvature
will be associated with the Yang-Mills fields
of the new kind defined on $CP(N-1)$. The unitary evolution being represented by the local dynamical variables 
$\Phi^i_\alpha, \Phi^i_\beta $ define the curvature 
in 2-dimension direction $(\alpha,\beta)$ 
\begin{eqnarray}
R(D_\alpha,D_\beta)X^k = 
[\nabla_{D_\alpha},\nabla_{D_\beta}] X^k -
\nabla_{[D_\alpha,D_\beta]} X^k \cr =
\{(D_\alpha \Phi^i_\beta - D_\beta \Phi^i_\alpha )
\Gamma^k_{in}  
-(\Phi^i_\beta \Phi^{*s}_\alpha -
\Phi^i_\alpha \Phi^{*s}_\beta) R^k_{i*sn} \cr -
 C^\gamma_{\alpha \beta} \Phi^i_\gamma \Gamma^k_{in}\}X^n. 
\label{R}  
\end{eqnarray}
Here $C^{\gamma}_{\alpha \beta}$ are of the $SU(N)$ group 
structure constants \cite{Le3}. 
Now we can introduce the follows expression for 
the curvature originated field 
$F_{\alpha \beta} = R (D_\alpha, D_\beta) X^k 
\frac{\partial}{\partial \pi^k} + c.c.$, 
which is the analog of Yang-Mills fields of the gauge 
potential associated with the intrinsic affine 
$CP(N-1)$ connection. To my mind the curvature of the coset manifold $CP(N-1)=SU(N)/S[U(1) \times U(N-1)]$ paves the way to the understanding the cooperative interaction in the ensemble of ``pseudo-spins" arose under the breakdown the dynamical group $G=SU(N)$ up to isotropy group 
$H=U(1) \times U(N-1)$. The equation for this field is not established up to now.

2. The second gauge potential realization is the truly parallel transport agrees with the Fubini-Study metric. 
Now I will introduce the parallel transported Hermitian dynamical variable $T^p$ obeys the follows equation
\begin{eqnarray}
\frac{d T^p}{ds}+\Gamma^p_{in}T^n\frac{d\pi^i}{ds}=0, c.c.,(s=\omega t).
\label{parallel}
\end{eqnarray}
This equation has the exact solutions along a geodesic. In the case $CP(1)$ it is as follows: $T^1(s)=(\xi (1+tan^2(\omega t))+ i\eta (1+tan^2(\omega t)))$. The scalar product $G_{ik^*}T^i(s) T^{k*}(s)=\xi^2+\eta^2$ is the invariant of the parallel transport.
Then the instant result of the comparison of the omnipresent along the geodesic the constant vector (initial conditions) $T^1(0)=(\xi+i\eta)$ and result of its parallel transport along the geodesic is as  follows
\begin{eqnarray}
\delta= G_{ik^*}T^i(0) T^{k*}(\omega t)=  (\xi^2+\eta^2)cos^2(\omega t).
\end{eqnarray}
The integral angle accumulated during the parallel transport of the dynamical variable along the geodesic is observable 
\begin{eqnarray}
\Lambda = \int_0^{\pi} arccos (\cos^2(\omega t)) dt 
\end{eqnarray}
as it depicted in the Fig.2.
\vskip .1cm
\begin{figure}[ht]
\centerline{\epsfxsize=\hsize \epsffile{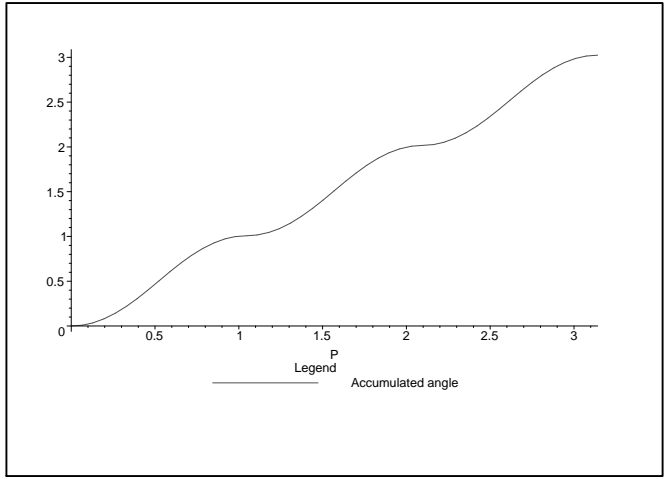}}
\caption{The observable accumulation of the difference between initial vector
and the result of the parallel transport along the geodesic. The  frequency of the dragging is as follows: $\omega=3$.}
\label{fig.2}
\end{figure} 

Let me to compare now the parallel transported vectors $T^1(s)$ and the vector 
$S^p=\Xi^p -\Gamma^p_{in} \Xi^n d \pi^i $  ``shifted" from $f(t)$ to the ``reference'' geodesic $\gamma(t)$ where $ d \pi^i = \pi^i(f(t))-\pi^i(\gamma(t))$. It means all local
tensors and $\Gamma^p_{in}$ were calculated on the ``reference'' geodesic. Result for the angle between these two vector along the ``reference" geodesic will be expressed through $\cos \chi$  
\begin{eqnarray}
\cos \chi (t)
= \frac{|G_{ik^*} T^i(\gamma(t)) S^{k*}(f(t))|}{||T|| ||S||}.
\end{eqnarray}
 
The result of the comparison of the parallel vectors fields arose from the initial vector $T^1(0)=(\xi+i \eta)$ parallel transported along the ``reference" geodesic $\gamma(t)$ and along the deformed geodesic $f(t)$ during the modulation is very interesting: {\it all vectors parallel transported along different paths looks like smoothly opening ``umbrella" along the geodesic. At the $\theta=\pi/2$ the parallel transported dynamical variable along one of the deformed geodesics $f(t)$ are orthogonal (in the sense of the Fubini-Study metric) to the ``handle" of the ``umbrella" - the parallel transported vector along the geodesic.
It means that in fact the result of the parallel transport is local: this is single-defined by the geodesic issued from the initial point and by the dynamical variable (tangent vector)}. 

The square of the cosine of the angle between the exact solution of the equation $T^i(\gamma(t))$ and the numerical solution
$\Xi^{k*}(f(t))$ for the parallel transport along deformed geodesic is shown in the picture Fig.3. 
\vskip .1cm
\begin{figure}[ht]
\centerline{\epsfxsize=\hsize \epsffile{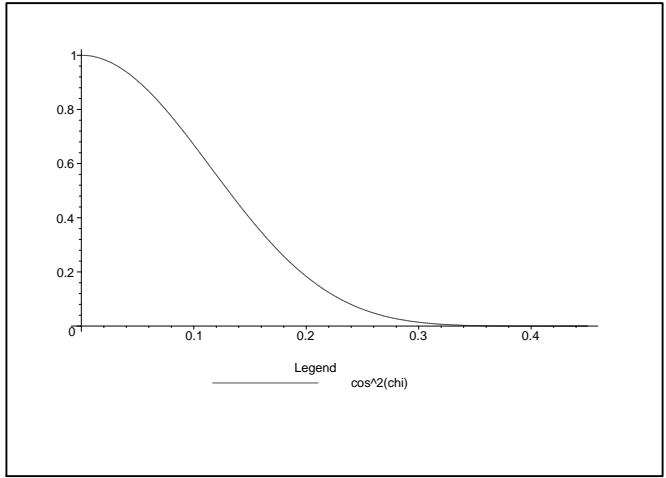}}
\caption{The square of cosine of the observable angle between two parallel transported vectors along the geodesic $\gamma(0,\pi/2)$ and deformed geodesic $f(0,\pi/2)$. The relationship between frequencies is as follows: $\Omega=10 \pi, \omega=\pi$.}
\label{fig.3}
\end{figure} 

{\bf Conjecture} 
The solution of the parallel transport equation along the geodesic
$\gamma(0,\pi/2)$ is orthogonal to any solution of the parallel transport equation along a path $f(0,\pi/2)$  at the $\theta=\pi/2$.

{\bf Comments}

1. If the foregoing conjecture is correct than we have the natural mechanism of the decoherence in $CP(1)$.

2. The covariant formulation of the quantum dynamics in the $CP(N-1)$ should lead to the observable geometrodynamical effects like discussed here path dependable behavior of the local dynamical variables. It may be verified by the measurement of the path dependent phase shift of the modulation frequency. On the other hand this approach may serves as a direct measurement of the sectional curvature of the projective Hilbert state space.

3. Local dynamical variables in $CP(1)$ gives a possibility to extract the  Qubit coordinates in the superposition state.
\vskip 2cm
 
{\bf Acknowledgments}

I sincerely thank Larry Horwitz for the interesting discussions of  non-linear modifications of quantum theory.

\end{document}